\documentclass[]{aa}
\usepackage{epsfig}
\begin{document}
\def\gsim{\vcenter{\hbox{$>$}\offinterlineskip\hbox{$\sim$}}}
\def\lsim{\vcenter{\hbox{$<$}\offinterlineskip\hbox{$\sim$}}}
\thesaurus{06(08.03.1; 08.03.4; 08.13.2; 08.16.4; 11.13.1; 13.09.6)}
\title{Mass-loss rates and luminosity functions of dust-enshrouded AGB stars
       and red supergiants in the LMC}
\author{Jacco Th. van Loon\inst{1}, M.A.T. Groenewegen\inst{3}, A. de
        Koter\inst{2}, Norman R. Trams\inst{4}, L.B.F.M. Waters\inst{2,5},
        Albert A. Zijlstra\inst{6}, Patricia A. Whitelock\inst{7}, Cecile
        Loup\inst{8}}
\institute{Institute of Astronomy, Madingley Road, Cambridge CB3 0HA, United
           Kingdom
      \and Astronomical Institute, University of Amsterdam, Kruislaan 403,
           NL-1098 SJ Amsterdam, The Netherlands
      \and Max-Planck Institut f\"{u}r Astrophysik, Karl-Schwarzschild
           Stra{\ss}e 1, D-85740 Garching bei M\"{u}nchen, Germany
      \and Astrophysics Division of ESA, ESTEC, P.O.Box 299, NL-2200 AG
           Noordwijk, The Netherlands
      \and Space Research Organization Netherlands, Landleven 12, NL-9700 AV
           Groningen, The Netherlands
      \and University of Manchester Institute of Science and Technology,
           P.O.Box 88, Manchester M60 1QD, United Kingdom
      \and South African Astronomical Observatory, P.O.Box 9, 7935
           Observatory, Republic of South Africa
      \and Institut d'Astrophysique de Paris, 98bis Boulevard Arago, F-75014
           Paris, France}
\offprints{Jacco van Loon, jacco@ast.cam.ac.uk}
\date{Received date; Accepted date}
\maketitle
\markboth{van Loon et al.: Dust-enshrouded AGB stars and RSGs in the LMC}
         {van Loon et al.: Dust-enshrouded AGB stars and RSGs in the LMC}
\begin{abstract}

A radiative transfer code is used to model the spectral energy distributions
of 57 mass-losing Asymptotic Giant Branch (AGB) stars and red supergiants
(RSGs) in the Large Magellanic Cloud (LMC) for which ISO spectroscopic and
photometric data are available. As a result we derive mass-loss rates and
bolometric luminosities.

A gap in the luminosity distribution around $M_{\rm bol} = -7.5$ mag separates
AGB stars from RSGs. The luminosity distributions of optically bright carbon
stars, dust-enshrouded carbon stars and dust-enshrouded M-type stars have only
little overlap, suggesting that the dust-enshrouded AGB stars are at the very
tip of the AGB and will not evolve significantly in luminosity before mass
loss ends their AGB evolution.

Derived mass-loss rates span a range from $\dot{M} \sim 10^{-7}$ to $10^{-3}$
M$_\odot$ yr$^{-1}$. More luminous and cooler stars are found to reach higher
mass-loss rates. The highest mass-loss rates exceed the classical limit set by
the momentum of the stellar radiation field, $L/c$, by a factor of a few due
to multiple scattering of photons in the circumstellar dust envelope.
Mass-loss rates are lower than the mass consumption rate by nuclear burning,
$\dot{M}_{\rm nuc}$, for most of the RSGs. Two RSGs have $\dot{M} \gg
\dot{M}_{\rm nuc}$, however, suggesting that RSGs shed most of their stellar
mantles in short phases of intense mass loss. Stars on the thermal pulsing AGB
may also experience episodes of intensified mass loss, but their quiescent
mass-loss rates are usually already higher than $\dot{M}_{\rm nuc}$.

\end{abstract}

\section{Introduction}

In their final stages of evolution, both intermediate-mass and more massive
stars become very large and assume very low photospheric temperatures. Either
as Asymptotic Giant Branch (AGB) stars for initial masses 1 M$_\odot
{\lsim}M<M_{\rm up}$ or as red supergiants (RSGs) for $M>M_{\rm up}$, with
$M_{\rm up}\sim8$ M$_\odot$, they become unstable and their mantles start to
pulsate with great amplitude. This presumably levitates matter out to a
distance from the star where gas temperature and gas density are favourable
for the formation of dust. Once dust forms in sufficient abundance, radiation
pressure on the dust grains and collisional coupling of the grains with the
gas drives an efficient stellar wind (Wickramasinghe et al.\ 1966; Goldreich
\& Scoville 1976). In this way, AGB stars and RSGs lose a significant fraction
of their initial mass at rates of up to $\dot{M} \sim 10^{-5}$ to $10^{-3}$
M$_\odot$ yr$^{-1}$, and they become surrounded by a dusty circumstellar
envelope (CSE). The stellar light is almost entirely absorbed by the dust at
wavelengths $\lambda\lsim1 \mu$m, but re-emission by the dust at
$\lambda\gsim10 \mu$m makes the CSEs very bright infrared (IR) objects. The
dust-enshrouded phase of evolution is particularly interesting for its
importance in the chemical enrichment of the interstellar medium (ISM) and the
formation of stellar remnants.

IRAS detected many IR point sources in the Large Magellanic Cloud (LMC)
(Schwering \& Israel 1990), from amongst which a few hundred candidates for
CSEs around AGB stars or RSGs could be selected (Reid et al.\ 1990; Loup et
al.\ 1997, hereafter paper I). Association with known optically bright stars
(mostly RSGs) and ground-based near-IR observations have resulted in the
identification of several dozen AGB star or RSG counterparts (Reid et al.\
1990; Wood et al.\ 1992; paper I; Zijlstra et al.\ 1996, hereafter paper II;
van Loon et al.\ 1997, hereafter paper III). Follow-up observations have been
primarily done to obtain accurate bolometric luminosities for these stars,
which is possible because the distance to the LMC is known with relatively
high accuracy, and also, in the case of AGB stars, to determine whether the
circumstellar dust and/or the photospheric composition is dominated by an
oxygenous or carbonaceous chemistry (van Loon et al.\ 1998, hereafter paper
IV; van Loon et al.\ 1999a). ISO spectroscopic and photometric observations of
a selection of 57 AGB stars and RSGs in the LMC were presented by Trams et
al.\ (1999b), including a thorough chemical classification study. In this
paper, the spectral energy distributions (SEDs) of the ISO sample are fitted
by a radiation transfer model, yielding mass-loss rates and bolometric
luminosities.

The paper is organised as follows: Section 2 describes the input to the model.
A number of assumptions have been made for various reasons, either because it
is a valid approximation, the parameter is not known, or simply for the sake
of restricting the parameter space. The aim has been to maximise homogeneity
in modelling our large dataset. Section 3 presents the observed SEDs and their
best model fits for all stars in our sample. In Section 4 the resulting
bolometric luminosities and mass-loss rates are discussed in detail.

\section{Input for the model fitting}

\subsection{The radiative transfer code}

To model the dusty CSE we have used a radiative transfer code (de Koter et
al., in preparation). It calculates the emergent SED from a star surrounded by
a spherically symmetric dusty CSE. Input parameters include the SED of the
underlying star, optical properties of the dust, and parameters describing the
CSE geometry. These are discussed below, after a brief description of the
observed SEDs.

\subsection{Observed spectral energy distributions}

The model is fitted to photometric and spectroscopic data in the IR. These
data consist of ISO measurements, supplemented by near-IR photometry gathered
at the South African Astronomical Observatory (SAAO) within the framework of a
monitoring campaign to obtain pulsation periods for the stars in the ISO
sample (Whitelock et al.\ 1999, in preparation). The variability information
resulting from the monitoring project will be used for correlation with the
results from the model fits. The ISO data comprise broadband photometry at 12,
25 and 60 $\mu$m using the ISOCAM and ISOPHOT instruments, and spectroscopy
between 2.5 and 12 $\mu$m using ISOPHOT and between 7 and 14 $\mu$m using
ISOCAM. The data are described and presented in detail by Trams et al.\
(1999b).

Interstellar extinction will be ignored, because the SED will only be studied
at wavelengths beyond 1 $\mu$m where interstellar extinction is expected to be
insignificant compared to the extinction by the CSE. However, there are
indications that along some lines of sight through the LMC this may not be
true (paper III). Furthermore, the distance to the stars is assumed to be 50
kpc. Although there is reason to believe that the distance to the LMC may be
somewhat larger than this (Feast 1998), most distance estimates cluster around
a value of 50 kpc (Walker 1999).

\subsection{Spectral energy distributions in the models}

Our sample of stars consist of (oxygen-rich) M-type stars and carbon stars. In
Trams et al.\ (1999b), and references therein, we classified the chemical
types of the dusty CSEs of most of the stars in our sample. For the stars for
which we do not have spectroscopic information about the stellar photosphere
we assume that the chemical composition of the dust corresponds to the carbon
or oxygen dominance in the stellar photosphere.

For M stars we use the synthetic spectra published by Fluks et al.\ (1994) for
the wavelength region between 0.1 and 12.5 $\mu$m, extended to 1 m by means of
a blackbody of the temperature given by Fluks et al. For carbon stars no such
data are readily available. Barnbaum et al.\ (1996) and L\'{a}zaro et al.\
(1994) present spectra of carbon stars at optical (0.4 to 0.7 $\mu$m) and
near-IR (1 to 4 $\mu$m) wavelengths, respectively. From their stars we
selected AQ Sgr, an N-type carbon star without infrared excess or strong
stellar pulsation, for use as a template.

The effective temperature $T_{\rm eff}$ of the stellar photosphere is only
known for a few optically bright stars in our sample where spectral subclasses
could be determined, ranging from M1 ($T_{\rm eff} = 3810$ K) to M10 ($T_{\rm
eff} = 2500$ K). For the oxygen-rich stars without this information we adopt
$T_{\rm eff} = 2890$ K, corresponding to a spectral type M8 III (Fluks et al.\
1994). Although cooler than the optically bright stars, this spectral type is
more appropriate for the oxygen-rich stars with higher mass-loss rates (see
paper IV). For the carbon star (AQ Sgr) we adopt $T_{\rm eff} = 2804$ K, as
measured by means of Lunar occultation techniques (L\'{a}zaro et al.\ 1994).
Stellar pulsation, a common feature of mass-losing AGB stars and RSGs, results
in an extended atmosphere, and the photospheric spectra for pulsating stars
are likely to be affected by this (Bowen 1988; Feast 1996). We ignore not only
these possible deviations from the input spectra that we use, but also the
variability of $T_{\rm eff}$ during the pulsation cycle. Finally, differences
in metallicity may affect the spectral classification and its relation to
effective temperature. Though important, it is a poorly explored subject.

\subsection{Dust properties}

%
%
\begin{figure}[tb]
\centerline{\psfig{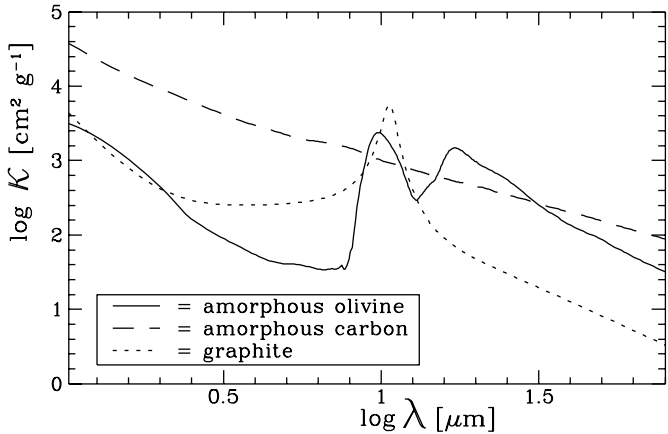}}
\caption[]{Opacities for the dust species used: amorphous olivine (solid;
J\"{a}ger et al.\ 1998), amorphous carbon (dashed; Preibisch et al.\ 1993) and
graphite (dotted; Laor \& Draine 1993).}
\end{figure}

The spectral coverage and resolution of the data do not enable the exact
composition and properties of the dust to be determined. Amorphous olivine and
carbon from the Jena group (J\"{a}ger et al.\ 1998; Preibisch et al.\ 1993)
are used for oxygen- and carbon-rich dust, respectively. Crystalline carbon in
the form of graphite (SiC), taken from Laor \& Draine 1993, is only added when
the SiC emission feature at $\sim11.3$ $\mu$m is observed. The opacities of
these three dust species are plotted in Fig.\ 1. Remaining discrepancies
between the best model fit and the observations will be discussed later, and
possible explanations include different (mixtures of) dust species. For
instance in oxygen-rich environments pyroxenes may contribute, as well as FeO,
and in some highly evolved CSEs crystalline silicates may be present (Waters
et al.\ 1999).

The spherical dust grains are assumed all to have the same radius $a = 0.1$
$\mu$m, and the optical properties of the grains are calculated assuming Mie
theory.

\subsection{Geometry of the circumstellar envelope}

The dusty CSE has two boundary conditions. The outer boundary is set by
insisting that the outer radius of the CSE is at $10^4$ stellar radii. This is
a rather arbitrary choice, but corresponds roughly to the typical extensions
of CSEs. An order of magnitude difference in this ratio does not significantly
affect the calculated SED. The inner boundary is set by requiring the
temperature at the inner radius, $R_{\rm in}$, of the dusty CSE to be $T_{\rm
dust} = 1000$ K. This corresponds to the typical temperature at which dust
condenses, although it should be realised that this depends on the chemistry.

The radiation transfer calculation requires the local density of dust,
${\rho}_{\rm dust}$, at a distance $r$ from the center of the star to be
specified. The continuity equation for a stationary outflow of a single fluid
prescribes
\begin{equation}
{\rho}_{\rm dust} = \frac{\dot{M}}{4 \pi \psi v_{\rm exp} r^2}
\end{equation}
with total (gas+dust) mass-loss rate $\dot{M}$, gas-to-dust (mass) ratio
$\psi$, and expansion velocity $v_{\rm exp}$. The parameter that we derive
from fitting the model to the observations is $\dot{M}$. This means that the
gas-to-dust ratio is assumed to be known. A value of $\psi = 500$ is adopted
here, believed to be representative of the moderately low LMC metallicity.
Little is known about the dependence of $\psi$ on the chemical composition of
the stellar atmosphere, the effective temperature and the density in the wind
at the location of dust condensation. Model calculations for carbon-rich
chemistry suggest that $\psi$ decreases with increasing amount of condensible
material and increases with increasing $T_{\rm eff}$ (Arndt et al.\ 1997).

Expansion velocities may be measured from the twin peaks of OH maser profiles,
but only a few of the brightest maser sources in the LMC are detectable with
current instrumentation (Wood et al.\ 1986, 1992; paper IV). A value of
$v_{\rm exp} = 10$ km s$^{-1}$ is often adopted for stars in the LMC. However,
$v_{\rm exp}$ depends on the stellar luminosity $L$ approximately as $v_{\rm
exp} \propto \sqrt[4]{L}$ (Jura 1984; Habing et al.\ 1994; see Arndt et al.\
1997 for a more detailed approximation). In our modelling we calibrate this
scaling relation by demanding a star with $L=30,000$ L$_\odot$ ($M_{\rm bol} =
-6.47$ mag) to have $v_{\rm exp} = 10$ km s$^{-1}$ (van Loon 1999). Another
complication is the fact that dust grains do not escape the stellar
gravitation field at a constant velocity. When dust forms in the proximity of
the star, matter in the stellar outflow is being accelerated. The velocity
$v_{\rm exp}$ as measured by OH is not reached instantaneously. However, dust
condensation is not complete instantaneously either, and the two effects tend
to cancel: in the dusty CSE the product $\psi(r) v_{\rm exp}(r)$, which
appears in Eq.\ (1), does not depend on $r$ (e.g.\ Groenewegen et al.\ 1998).

The radial optical depth $\tau_{\rm rad}$ is the optical depth along the
line-of-sight towards the star, and determines the amount of extinction and
the temperature structure throughout the dusty CSE. The maximal tangential
optical depth $\tau_{\rm tan}$ is the optical depth along the line-of-sight
parallel to the line-of-sight towards the star, at a minimal distance $R_{\rm
in}$ to the star, and measures the amount of emission from the dusty CSE. It
is assumed here that ${\rho}_{\rm dust} \propto r^{-2}$, i.e.\ $\dot{M}$ is
constant in time. For this simple geometry, with $R_{\rm in} \ll R_{\rm out}$,
the radial and tangential optical depths are related as
\begin{equation}
\tau_{\rm tan} / \tau_{\rm rad} = \pi
\end{equation}
When presenting the results of the model fits, the values of $\tau_{\rm rad}$
at a wavelength $\lambda=1 \mu$m are listed.

\subsection{Fitting strategy}

%
%
\begin{table}
\caption[]{Model parameters for M-type stars with ISO spectra, using amorphous
olivine from Jena. For a given star, $R_\star$ (in $R_\odot$), $R_{\rm in}$
(in $R_\star$), and $\tau_{\rm rad}$ (at a wavelength $\lambda=1 \mu$m) are
adjusted to fit the model to the observations, yielding $\dot{M}$ (in
$10^{-6}$ M$_\odot$ yr$^{-1}$). Model fits derived for multiple epochs are
labelled in the last column.}
\begin{tabular}{llrllrl}
\hline\hline
\vspace*{-3mm}\\
Star                           &
$T_{\rm eff}$                  &
$R_\star$                      &
$R_{\rm in}$                   &
$\tau_{\rm rad}$               &
$\dot{M}$                      &
Ep\rlap{.}                     \\
\hline
HV12070                        &
3309                           &
600                            &
\llap{1}1.2                    &
0.021                          &
0\rlap{.5}                     &
                               \\
HV2446                         &
3434                           &
540                            &
\llap{1}3.3                    &
0.054                          &
1\rlap{.5}                     &
                               \\
HV888                          &
3574                           &
\llap{1}300                    &
\llap{1}5.5                    &
0.041                          &
5                              &
                               \\
HV996                          &
3574                           &
950                            &
\llap{1}6.5                    &
0.16                           &
13                             &
                               \\
IRAS04545$-$7000               &
2890                           &
850                            &
\llap{1}2.0                    &
7.1                            &
280                            &
                               \\
IRAS05003$-$6712               &
2890                           &
560                            &
9.5                            &
1.7                            &
29                             &
                               \\
IRAS05298$-$6957               &
2890                           &
400                            &
\llap{1}3.0                    &
\llap{1}4                      &
230                            &
A                              \\
                               &
2890                           &
800                            &
\llap{1}2.0                    &
7.5                            &
230                            &
B                              \\
IRAS05329$-$6708               &
2890                           &
600                            &
\llap{1}2.0                    &
7.7                            &
180                            &
                               \\
IRAS05402$-$6956               &
2890                           &
850                            &
\llap{1}1.5                    &
5.1                            &
180                            &
A                              \\
                               &
2890                           &
550                            &
\llap{1}2.0                    &
7.6                            &
180                            &
B                              \\
IRAS05558$-$7000               &
2890                           &
750                            &
9.8                            &
2.1                            &
50                             &
A                              \\
                               &
2890                           &
480                            &
\llap{1}0.5                    &
3.0                            &
50                             &
B                              \\
SP77 30$-$6                    &
2890                           &
700                            &
8.0                            &
0.35                           &
7                              &
                               \\
WOH G64                        &
3126                           &
\llap{2}100                    &
\llap{1}6.0                    &
\llap{1}2                      &
\llap{2}800                    &
                               \\
\hline
\end{tabular}
\end{table}

%
%
\begin{table}
\caption[]{Model parameters for M-type stars without ISO spectra, using
amorphous olivine from Jena. For a given star, $R_\star$ (in $R_\odot$),
$R_{\rm in}$ (in $R_\star$), and $\tau_{\rm rad}$ (at a wavelength $\lambda=1
\mu$m) are adjusted to fit the model to the observations, yielding $\dot{M}$
(in $10^{-6}$ M$_\odot$ yr$^{-1}$).}
\begin{tabular}{llrllr}
\hline\hline
\vspace*{-3mm}\\
Star                           &
$T_{\rm eff}$                  &
$R_\star$                      &
$R_{\rm in}$                   &
$\tau_{\rm rad}$               &
$\dot{M}$                      \\
\hline
HV12501                        &
3810                           &
840                            &
\llap{1}9.7                    &
0.034                          &
3                              \\
HV2360                         &
3736                           &
870                            &
\llap{1}9.2                    &
0.083                          &
7                              \\
HV5870                         &
3434                           &
820                            &
\llap{1}3.6                    &
0.10                           &
5                              \\
HV916                          &
3666                           &
1050                           &
\llap{1}7.5                    &
0.060                          &
6                              \\
IRAS04407$-$7000               &
2890                           &
1150                           &
8.9                            &
1.1                            &
50                             \\
IRAS04498$-$6842               &
2890                           &
770                            &
8.9                            &
1.0                            &
26                             \\
IRAS04509$-$6922               &
2500                           &
950                            &
6.0                            &
0.83                           &
17                             \\
IRAS04516$-$6902               &
2667                           &
1200                           &
7.0                            &
0.93                           &
33                             \\
IRAS04530$-$6916               &
2890                           &
1400                           &
\llap{1}2.5                    &
9.5                            &
840                            \\
IRAS05294$-$7104               &
2890                           &
800                            &
9.9                            &
2.1                            &
64                             \\
SHV0522023$-$701242            &
3666                           &
160                            &
\llap{1}6.7                    &
0.001                          &
$\lsim$0\rlap{.01}             \\
SHV0524565$-$694559            &
3434                           &
190                            &
\llap{1}3.1                    &
0.034                          &
0\rlap{.2}                     \\
SHV0530323$-$702216            &
3309                           &
280                            &
\llap{1}1.3                    &
0.053                          &
0\rlap{.4}                     \\
WOH SG374                      &
3309                           &
730                            &
\llap{1}3.0                    &
0.59                           &
23                             \\
\hline
\end{tabular}
\end{table}

%
%
\begin{table}
\caption[]{Model parameters for carbon stars with ISO spectra (assuming type
C5 with $T_{\rm eff} = 2804$ K), using amorphous carbon from Jena and
crystalline carbon (SiC) from Laor \& Draine (1993). For a given star,
$R_\star$ (in $R_\odot$), $R_{\rm in}$ (in $R_\star$), the mass fraction (in
\%) of dust contained into SiC, and $\tau_{\rm rad}$ (at a wavelength
$\lambda=1 \mu$m) are adjusted to fit the model to the observations, yielding
$\dot{M}$ (in $10^{-6}$ M$_\odot$ yr$^{-1}$). Model fits derived for multiple
epochs are labelled in the last column.}
\begin{tabular}{lrlrlrl}
\hline\hline
\vspace*{-3mm}\\
Star                           &
$R_\star$                      &
$R_{\rm in}$                   &
SiC                            &
$\tau_{\rm rad}$               &
$\dot{M}$                      &
Ep\rlap{.}                     \\
\hline
HV2379                         &
330                            &
9.5                            &
0                              &
1.4                            &
0\rlap{.6}                     &
A                              \\
                               &
270                            &
9.6                            &
0                              &
1.7                            &
0\rlap{.6}                     &
B                              \\
IRAS04374$-$6831               &
480                            &
\llap{1}0.6                    &
0                              &
7.4                            &
8                              &
                               \\
IRAS05112$-$6755               &
580                            &
\llap{1}0.8                    &
0                              &
\llap{1}1                      &
14                             &
                               \\
IRAS05128$-$6455               &
600                            &
\llap{1}0.3                    &
0                              &
5.0                            &
6                              &
                               \\
IRAS05190$-$6748               &
520                            &
\llap{1}1.0                    &
0                              &
\llap{1}7                      &
19                             &
                               \\
IRAS05289$-$6617               &
290                            &
\llap{1}2.6                    &
20                             &
\llap{6}7                      &
41                             &
A                              \\
                               &
220                            &
\llap{1}3.3                    &
20                             &
\llap{8}4                      &
41                             &
B                              \\
IRAS05348$-$7024               &
650                            &
\llap{1}1.1                    &
50                             &
\llap{1}2                      &
26                             &
A                              \\
                               &
500                            &
\llap{1}1.3                    &
50                             &
9.2                            &
26                             &
B                              \\
IRAS05568$-$6753               &
450                            &
\llap{1}2.2                    &
0                              &
\llap{5}6                      &
55                             &
                               \\
SHV0500193$-$681706            &
340                            &
\llap{1}0.0                    &
0                              &
2.2                            &
1\rlap{.9}                     &
A                              \\
                               &
500                            &
9.8                            &
0                              &
3.2                            &
1\rlap{.9}                     &
B                              \\
SHV0500233$-$682914            &
480                            &
9.7                            &
20                             &
1.5                            &
1\rlap{.6}                     &
                               \\
\hline
\end{tabular}
\end{table}

%
%
\begin{table}
\caption[]{Model parameters for carbon stars without ISO spectra (assuming
type C5 with $T_{\rm eff} = 2804$ K), using amorphous carbon from Jena. For a
given star, $R_\star$ (in $R_\odot$), $R_{\rm in}$ (in $R_\star$), and
$\tau_{\rm rad}$ (at a wavelength $\lambda=1 \mu$m) are adjusted to fit the
model to the observations.}
\begin{tabular}{lrllr}
\hline\hline
\vspace*{-3mm}\\
Star                           &
$R_\star$                      &
$R_{\rm in}$                   &
$\tau_{\rm rad}$               &
$\dot{M}$                      \\
\hline
GRV0519$-$6700                 &
320                            &
8.5                            &
0.016                          &
$\lsim$0\rlap{.01}             \\
IRAS04286$-$6937               &
430                            &
\llap{1}0.4                    &
6.4                            &
5                              \\
IRAS04539$-$6821               &
490                            &
\llap{1}0.7                    &
\llap{1}0                      &
10                             \\
IRAS04557$-$6753               &
410                            &
\llap{1}0.8                    &
\llap{1}1                      &
8                              \\
IRAS05009$-$6616               &
570                            &
\llap{1}0.4                    &
6.7                            &
8                              \\
IRAS05113$-$6739               &
520                            &
\llap{1}0.8                    &
\llap{1}0                      &
11                             \\
IRAS05295$-$7121               &
410                            &
\llap{1}0.6                    &
8.1                            &
6                              \\
IRAS05300$-$6651               &
480                            &
\llap{1}0.8                    &
\llap{1}2                      &
11                             \\
IRAS05360$-$6648               &
430                            &
\llap{1}0.8                    &
\llap{1}1                      &
9                              \\
IRAS05506$-$7053               &
480                            &
\llap{1}1.0                    &
\llap{1}7                      &
16                             \\
SHV0454030$-$675031            &
190                            &
9.2                            &
0.76                           &
0\rlap{.15}                    \\
SHV0502469$-$692418            &
290                            &
8.5                            &
0.018                          &
$\lsim$0\rlap{.01}             \\
SHV0521050$-$690415            &
680                            &
8.9                            &
0.44                           &
0\rlap{.6}                     \\
SHV0522118$-$702517            &
300                            &
9.6                            &
1.7                            &
0\rlap{.7}                     \\
SHV0526001$-$701142            &
360                            &
9.5                            &
1.5                            &
0\rlap{.8}                     \\
SHV0535442$-$702433            &
320                            &
8.8                            &
0.23                           &
0\rlap{.1}                     \\
TRM45                          &
350                            &
\llap{1}0.4                    &
6.0                            &
3\rlap{.5}                     \\
TRM72                          &
460                            &
\llap{1}0.3                    &
5.6                            &
5                              \\
TRM88                          &
490                            &
\llap{1}0.0                    &
3.1                            &
3                              \\
WBP14                          &
330                            &
9.3                            &
1.0                            &
0\rlap{.5}                     \\
\hline
\end{tabular}
\end{table}

The strategy to obtain the best fit of the model to the observations involves
iterations of model calculations where input parameters are being adjusted
according to the following (rough) guidelines: (i) adjust $R_\star$ to
reproduce $L$; (ii) adjust $\dot{M}$ to fit the 25 $\mu$m flux density; (iii)
adjust $R_{\rm inner}$ to fit the extinction in the near-IR. The aim is to
reach $T_{\rm dust} = 1000$ K at $R_{\rm in}$, where $T_{\rm dust}$ refers to
the hottest component in the case of a mixture of dust species. If a spectrum
around 10 $\mu$m is available, then the strength of the silicate (or SiC)
feature and the level of the stellar and/or dust continuum are additional
criteria to be met. First $v_{\rm exp} = 10$ km s$^{-1}$ is used, but
afterwards $v_{\rm exp}$ and $\dot{M}$ are scaled as described in Section 2.5,
using $L$ from the final fit of the model to the observations. The resulting
values for $L$ and $\dot{M}$ --- as opposed to $R_{\rm inner}$ --- are rather
insensitive to the exact value of $T_{\rm eff}$, which for many of our stars
is not known from observations.

\section{Results of model fitting}

%
%
\begin{figure*}[tb]
\centerline{\psfig{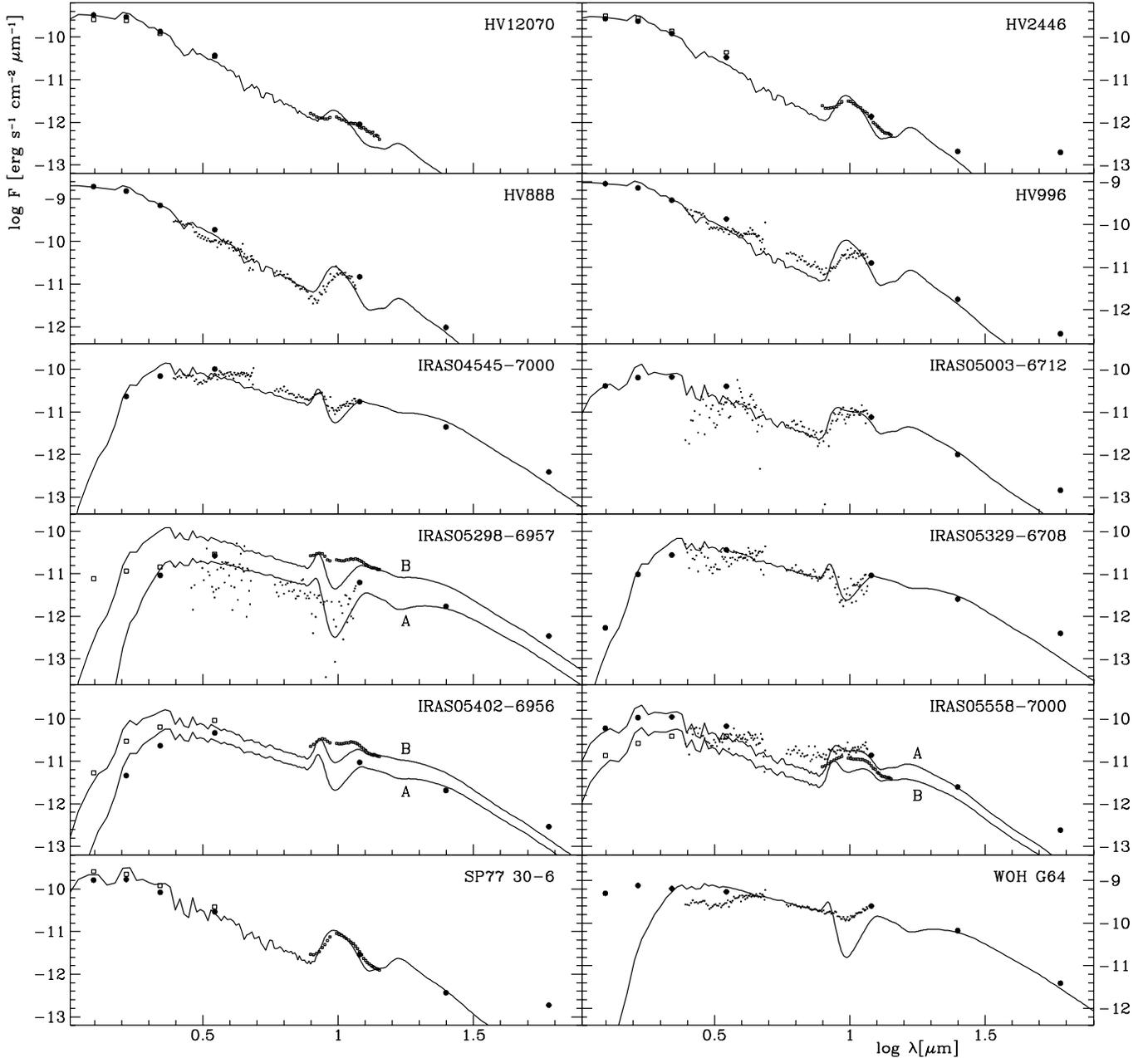}}
\caption[]{Model fits (solid lines, Table 1) and observational data (solid
circles for epoch A and open squares for epoch B) for M-type stars with ISO
spectra.}
\end{figure*}

The results of the model fits are listed in Tables 1 to 5, and the model SEDs
are presented together with the observational data in Figs.\ 2 to 6. The stars
are discussed in separate sections according to the chemistry and the
availability of ISO spectroscopy. The peculiar carbon star IRAS04496$-$6958 is
described in a section of its own. When more than one epoch of data was
available, both epochs were fitted with the requirement that the mass-loss
rate remains the same. The approach taken here is not entirely correct, as the
stars are all Long Period Variables requiring dynamical models (Hron et al.\
1998).

\subsection{M-type stars with ISO spectra}

The input spectral types and model fit parameters for the M-type stars for
which we have ISO spectroscopic data are listed in Table 1, and the model fits
are plotted together with the photometric and spectroscopic data in Fig.\ 2.

%
%
\begin{figure*}[tb]
\centerline{\psfig{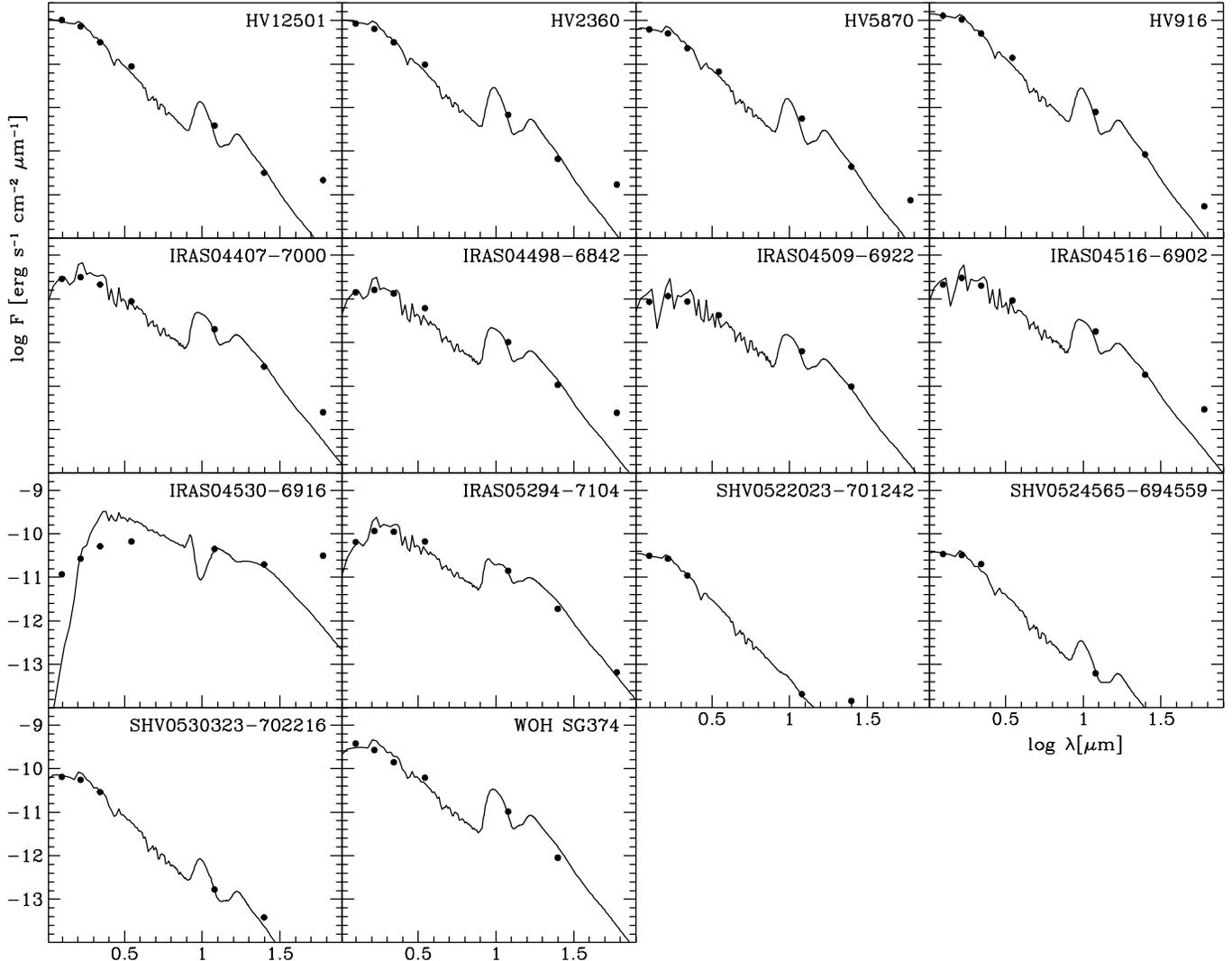}}
\caption[]{Model fits (solid lines, Table 2) and observational data for M-type
stars without ISO spectra.}
\end{figure*}

Examples of good fits are IRAS05329$-$6708 and SP77 30$-$6, for which both the
general shape of the SED and the shape and strength of the 10 $\mu$m silicate
feature are well reproduced by the model. The only serious flaw in the fit for
these stars is the discrepancy between the observed and predicted 60 $\mu$m
flux density. This is seen in all fits, without exception, including carbon
stars. The data do not reveal any trend with luminosity, mass-loss rate or IR
colour, nor with the time that a dust particle spends in the dust shell, $t =
10^4 R_\star / V_{\rm exp}$, or the density at the inner radius of the dust
shell, $\rho = \dot{M} / 4 \pi R_{\rm in}^2 v_{\rm exp}$. There are three
possible explanations: (i) the measured 60 $\mu$m flux densities may have been
over-estimated (Trams et al.\ 1999b); (ii) a contribution of large ($a\gsim1
\mu$m) and/or cool ($T\lsim200$ K) grains would enhance the emission at far-IR
wavelengths, in which case a trend of increasing $F_{\rm observed}/F_{\rm
model}$ [60 $\mu$m] for larger $t$ or $\rho$ is expected. This cannot be
confirmed with the current dataset, but the quality of the 60 $\mu$m data is
not high enough to rule it out either; (iii) a shallower slope of the SED at
far-IR wavelengths may be obtained by increasing the complex term of the
optical constants, thereby increasing the emission in the far-IR over the
extinction at shorter wavelengths.

Often the silicate feature produced by the model using amorphous olivine is
too strong compared to that observed, i.e.\ for HV996 and IRAS05402$-$6956.
For HV12070 this may be due to the fact that it is an MS-type star, i.e.\ a
star with a carbon-to-oxygen ratio smaller than but close to 1, and hence the
dust may be of a rather peculiar chemical composition. The observed silicate
emission sometimes peaks at a longer wavelength than does the model, which may
be caused by our assumption of spherical grain shape: a continuous
distribution of ellipsoids would yield somewhat broader emission features at a
wavelength that can be a few 0.1 $\mu$m longer. Perhaps the silicate dust in
the LMC is different from silicate dust that is formed at solar metallicity.

A third discrepancy seen for some, but not all, stars is an emission
deficiency in the model fit in the wavelength interval between $\sim4$ and 8
$\mu$m, i.e.\ for HV996 and IRAS05558$-$7000. This is known from the
literature to be a common problem in fitting models to oxygen-rich CSEs, and
it is investigated in detail for the post-AGB star AFGL4106 by Molster et al.\
(1999). Possible explanations include small, hot grains, or a different
chemical composition of the dust.

%
%
\begin{figure*}[tb]
\centerline{\psfig{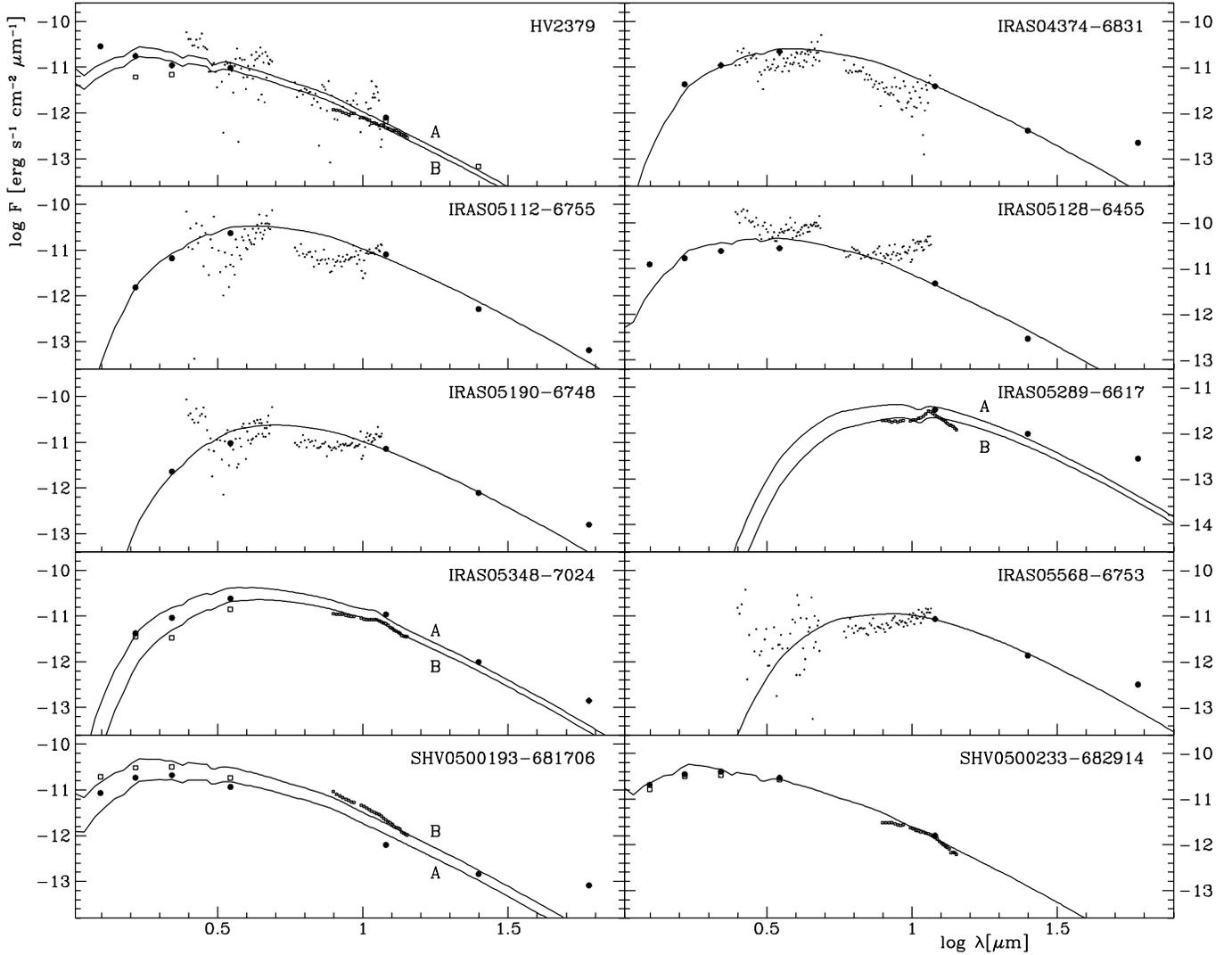}}
\caption[]{Model fits (solid lines, Table 3) and observational data (solid
circles for epoch A and open squares for epoch B) for carbon stars with ISO
spectra.}
\end{figure*}

The two stars for which we had most difficulties in reaching an acceptable fit
are IRAS05298$-$6957 and WOH G64. The common problem is reproducing the
observed weak extinction in the near-IR and the silicate feature compared to
the strong mid- and far-IR emission. One possible explanation is that of a
disk-like shape of the CSE, in which case the disks of these stars are viewed
under a large inclination angle causing little extinction in the line-of-sight
towards the stars. Perhaps we see here the progenitors of the disk-like and
bipolar morphologies common for planetary nebulae and post-RSGs.
Alternatively, their dusty CSEs may be clumpy (see e.g.\ the recent spatially
resolved observations of IRC+10216 by Weigelt et al.\ 1998 and AFGL 4106 by
van Loon et al.\ 1999b).

\subsection{M-type stars without ISO spectra}

The input spectral types and model fit parameters for the M-type stars for
which we do not have ISO spectroscopic data are listed in Table 2, and the
model fits are plotted together with the photometric data in Fig.\ 3.

In general, good fits of the model to the observed SEDs are obtained,
disregarding the 60 $\mu$m flux densities. This is likely connected to the
fact that there is no information about the strength of the silicate feature
or the emission between 4 and 8 $\mu$m. IRAS04530$-$6916 may well be an
example of a star like WOH G64, for which we argued that the CSE may be
disk-like.

The only M-type star that has an SED consistent with no mass loss is
SHV0522023$-$701242 (the ISOPHOT measurement at 25 $\mu$m for this star is
consistent with no detection), indicating that we are able to detect mass-loss
rates in excess of a few $\times 10^{-8}$ M$_\odot$ yr$^{-1}$.

%
%
\begin{figure*}[tb]
\centerline{\psfig{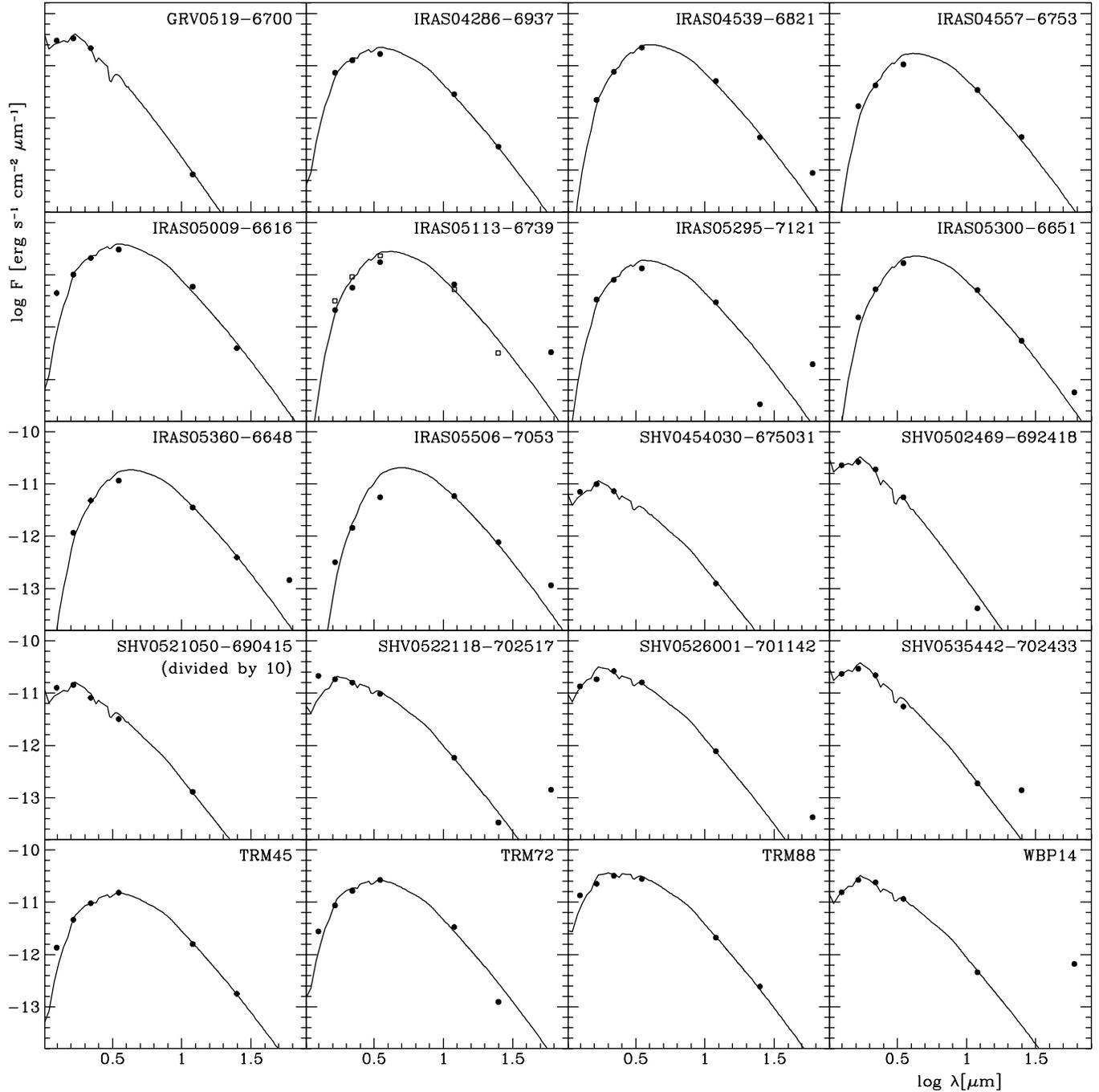}}
\caption[]{Model fits (solid lines, Table 4) and observational data (solid
circles for epoch A and open squares for epoch B) for carbon stars without ISO
spectra. The 12 and 25 $\mu$m flux densities of IRAS05506$-$7053 are IRAS data
(Trams et al.\ 1999b).}
\end{figure*}

\subsection{Carbon stars with ISO spectra}

The model fit parameters for the carbon stars for which we have ISO
spectroscopic data are listed in Table 3, and the model fits are plotted
together with the photometric and spectroscopic data in Fig.\ 4.

The general shape of the SED can be reproduced by the model rather well, but
in several cases the observed spectra lack emission in the wavelength interval
between $\sim5$ and 10 $\mu$m. This may be attributed to the absorption by
carbonaceous molecules (Hron et al.\ 1998). The discrepancy between the ISO
spectrum and the photometric data of IRAS05128$-$6455 may be related to a
problem in the background subtraction of the PHOT-S spectrum.

%
%
\begin{figure*}[tb]
\centerline{\psfig{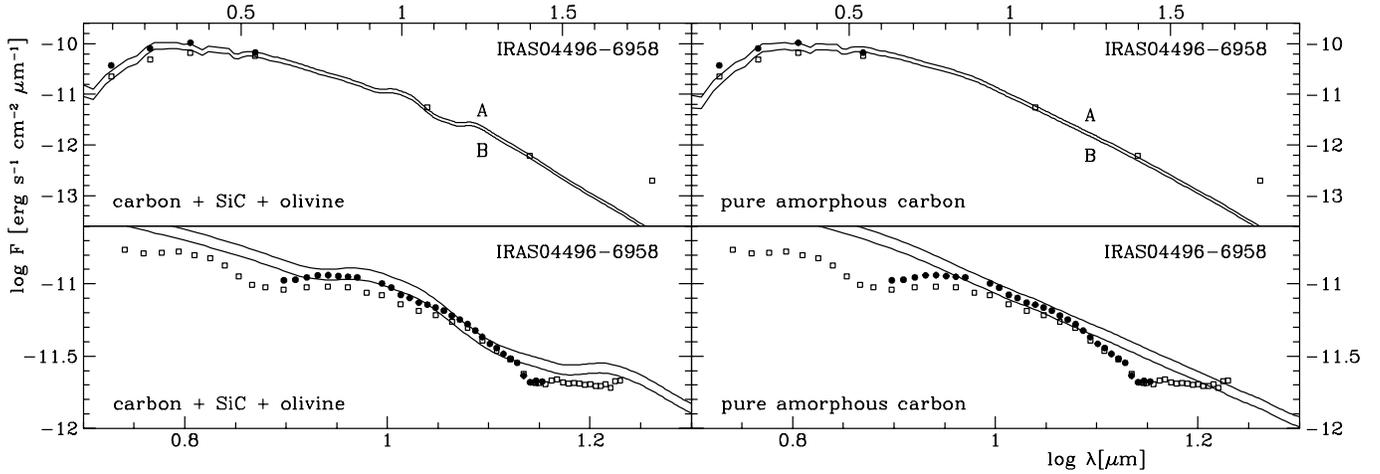}}
\caption[]{Model fit to the observed spectral energy distribution of the
peculiar carbon star IRAS04496$-$6958 (Table 5), using a mixture of amorphous
carbon, silicon carbide and amorphous olivine (left), and using pure amorphous
carbon (right).}
\end{figure*}

The strength of the silicon carbide (SiC) feature in the spectrum of
IRAS05348$-$7024 could be well reproduced by a mixture of equal quantities of
amorphous carbon and crystalline SiC. This would imply a very high crystalline
dust content. The SiC fractions are sensitive to the optical constants used
--- which may be metallicity dependent --- and to possible differences in the
geometrical distributions of the dust species. Such flaws in our modelling
might also explain the problem in reproducing the SiC feature in the very
obscured object IRAS05289$-$6617.

\subsection{Carbon stars without ISO spectra}

The model fit parameters for the carbon stars for which we do not have ISO
spectroscopic data are listed in Table 4, and the model fits are plotted
together with the photometric data in Fig.\ 5. The ISOPHOT photometry of
IRAS05506$-$7053 may have been in error, the 12 and 25 $\mu$m flux densities
being consistent with no detection whilst even without an IR excess this star
should have been detected easily. Adopting the IRAS photometry the SED may be
(poorly) fitted using amorphous carbon instead of silicates. Hence we have
reclassified this star as a carbon star, superseding the result obtained by
Trams et al.\ (1999b).

All SEDs can be fit satisfactorily, disregarding the 60 $\mu$m flux densities.
The ISOPHOT measurement at 25 $\mu$m of IRAS05295$-$7121, which is at a very
faint level indeed, bears a large uncertainty and is within the errorbars
consistent with the model fit.

Two stars, GRV0519$-$6700 and SHV0502469$-$692418, have SEDs consistent with
no mass loss: like for oxygen-rich CSEs, mass-loss rates in excess of a few
$\times 10^{-8}$ M$_\odot$ yr$^{-1}$ are detectable in our data.

\subsection{IRAS04496$-$6958}

The ISO data of this star have been discussed extensively in Trams et al.\
(1999a), where the authors argue for the presence of a minor fraction of
silicate dust in the CSE.

Here we fit the SED of this silicate carbon star with a dust mixture of
amorphous carbon, olivine and SiC, and also using pure amorphous carbon dust.
The model fit parameters for the carbon star IRAS04496$-$6958 are listed in
Table 5, and the model fits are plotted together with the photometric and
spectroscopic data in Fig.\ 6.

%
%
\begin{table}
\caption[]{Model parameters for the silicate carbon star IRAS04496$-$6958,
assuming spectral type C5 with $T_{\rm eff} = 2804$ K, and using amorphous
carbon and olivine from Jena and crystalline carbon (SiC) from Laor \& Draine
(1993). $R_\star$ (in $R_\odot$), $R_{\rm in}$ (in $R_\star$), the mass
fraction (in \%) of dust contained into SiC and olivine, respectively, and
$\tau_{\rm rad}$ (at a wavelength $\lambda=1 \mu$m are adjusted to fit the
model to the observations, yielding $\dot{M}$ (in $10^{-6}$ M$_\odot$
yr$^{-1}$). Model fits derived for multiple epochs are labelled in the last
column}
\begin{tabular}{lrlrrlrl}
\hline\hline
\vspace*{-3mm}\\
Model                          &
$R_\star$                      &
$R_{\rm in}$                   &
SiC                            &
olivine                        &
$\tau_{\rm rad}$               &
$\dot{M}$                      &
Ep\rlap{.}                     \\
\hline
Sil.+carbon                    &
760                            &
\llap{1}0.0                    &
10                             &
60                             &
2.4                            &
12                             &
A                              \\
                               &
680                            &
\llap{1}0.1                    &
10                             &
60                             &
2.6                            &
12                             &
B                              \\
Carbon                         &
820                            &
9.9                            &
0                              &
0                              &
3.0                            &
5\rlap{.6}                     &
A                              \\
                               &
730                            &
\llap{1}0.1                    &
0                              &
0                              &
3.3                            &
5\rlap{.6}                     &
B                              \\
\hline
\end{tabular}
\end{table}

Because of the smaller opacity of olivine, the mixture requires lower
luminosities and higher mass-loss rates in order to fit the observed SED than
in the case of pure carbon only. The shape of the SED around 10 $\mu$m is
partly reproduced using the mixture, featuring emission from olivine at
$\sim10$ and $18 \mu$m, and from SiC at $\sim11.5 \mu$m. The oxygen-rich dust
component is likely to be in a stationary disk configuration rather than
evenly distributed in the spherically symmetric, outflowing CSE (Trams et al.\
1999a). This may be the cause for the very high estimate of the olivine
fraction of 60\%, which disagrees with the observed IR colours that are
entirely dominated by carbon-rich dust. Hence we adopt instead the luminosity
and mass-loss rate derived when using pure carbon. The model fit using pure
carbon would imply strong molecular absorption in the 5 to 9 $\mu$m region and
around 14 $\mu$m (see also Hron et al.\ 1998).

%
%
\begin{table}
\caption[]{Bolometric magnitudes $M_{\rm bol}$, absolute K-band magnitudes
$M_{\rm K}$, $(K-L)$ colours and bolometric corrections $BC_{\rm K}$ relative
to $M_{\rm K}$ of the M-type stars. The distance to the LMC is assumed $d =
50$ kpc, and the bolometric magnitude of the Sun $M_{{\rm bol},\odot} =
-4.72$.}
\begin{tabular}{lrrcr}
\hline\hline
Star                           &
$M_{\rm bol}$                  &
$M_{\rm K}$                    &
$(K-L)$                        &
$BC_{\rm K}$                   \\
\hline
HV12070                        &
$-$6.75                        &
$-$9.75                        &
0.47                           &
3.00                           \\
HV12501                        &
$-$8.09                        &
$-$10.74                       &
0.45                           &
2.65                           \\
HV2360                         &
$-$8.09                        &
$-$10.74                       &
0.55                           &
2.65                           \\
HV2446                         &
$-$6.69                        &
$-$9.75                        &
0.53                           &
3.06                           \\
HV5870                         &
$-$7.60                        &
$-$10.39                       &
0.50                           &
2.79                           \\
HV888                          &
$-$8.77                        &
$-$11.60                       &
0.43                           &
2.83                           \\
HV916                          &
$-$8.42                        &
$-$11.24                       &
0.45                           &
2.82                           \\
HV996                          &
$-$8.09                        &
$-$10.91                       &
0.75                           &
2.82                           \\
IRAS04407$-$7000               &
$-$7.52                        &
$-$10.31                       &
0.88                           &
2.79                           \\
IRAS04498$-$6842               &
$-$6.65                        &
$-$9.79                        &
1.00                           &
3.14                           \\
IRAS04509$-$6922               &
$-$6.42                        &
$-$9.34                        &
1.05                           &
2.92                           \\
IRAS04516$-$6902               &
$-$7.22                        &
$-$10.25                       &
1.01                           &
3.03                           \\
IRAS04530$-$6916               &
$-$7.93                        &
$-$8.76                        &
2.13                           &
0.83                           \\
IRAS04545$-$7000               &
$-$6.85                        &
$-$9.09                        &
2.25                           &
2.24                           \\
IRAS05003$-$6712               &
$-$5.95                        &
$-$9.04                        &
1.30                           &
3.09                           \\
IRAS05294$-$7104               &
$-$6.72                        &
$-$9.59                        &
1.30                           &
2.87                           \\
IRAS05298$-$6957               &
$-$5.21                        &
$-$6.89                        &
3.00                           &
1.68                           \\
                               &
$-$6.72                        &
$-$7.39                        &
2.60                           &
0.67                           \\
IRAS05329$-$6708               &
$-$6.09                        &
$-$8.09                        &
2.15                           &
2.00                           \\
IRAS05402$-$6956               &
$-$5.90                        &
$-$7.89                        &
2.60                           &
1.99                           \\
                               &
$-$6.85                        &
$-$8.99                        &
2.25                           &
2.14                           \\
IRAS05558$-$7000               &
$-$6.58                        &
$-$9.59                        &
1.30                           &
3.01                           \\
                               &
$-$5.61                        &
$-$8.46                        &
1.83                           &
2.85                           \\
SHV0522023$-$701242            &
$-$4.33                        &
$-$7.09                        &
                               &
2.76                           \\
SHV0524565$-$694559            &
$-$4.42                        &
$-$7.74                        &
                               &
3.32                           \\
SHV0530323$-$702216            &
$-$5.10                        &
$-$8.14                        &
                               &
3.04                           \\
SP77 30$-$6                    &
$-$6.45                        &
$-$9.49                        &
0.65                           &
3.04                           \\
WOH G64                        &
$-$9.19                        &
$-$11.51                       &
1.66                           &
2.32                           \\
WOH SG374                      &
$-$7.17                        &
$-$9.85                        &
0.95                           &
2.68                           \\
\hline
\end{tabular}
\end{table}

%
%
\begin{table}
\caption[]{Bolometric magnitudes $M_{\rm bol}$, absolute K-band magnitudes
$M_{\rm K}$, $(K-L)$ colours, and bolometric corrections $BC_{\rm K}$ relative
to $M_{\rm K}$ of the carbon stars. The distance to the LMC is assumed $d =
50$ kpc, and the bolometric magnitude of the Sun $M_{{\rm bol},\odot} =
-4.72$.}
\begin{tabular}{lrrcr}
\hline\hline
Star                           &
$M_{\rm bol}$                  &
$M_{\rm K}$                    &
$(K-L)$                        &
$BC_{\rm K}$                   \\
\hline
GRV0519$-$6700                 &
$-$4.68                        &
$-$7.82                        &
                               &
3.14                           \\
HV2379                         &
$-$4.73                        &
$-$7.09                        &
1.70                           &
2.36                           \\
                               &
$-$4.29                        &
$-$6.59                        &
                               &
2.30                           \\
IRAS04286$-$6937               &
$-$5.30                        &
$-$7.24                        &
2.15                           &
1.96                           \\
IRAS04374$-$6831               &
$-$5.54                        &
$-$7.09                        &
2.55                           &
1.55                           \\
IRAS04496$-$6958               &
$-$6.70                        &
$-$9.54                        &
1.35                           &
2.84                           \\
                               &
$-$6.45                        &
$-$9.04                        &
1.60                           &
2.59                           \\
IRAS04539$-$6821               &
$-$5.59                        &
$-$6.69                        &
3.00                           &
1.00                           \\
IRAS04557$-$6753               &
$-$5.20                        &
$-$6.04                        &
2.85                           &
0.84                           \\
IRAS05009$-$6616               &
$-$5.92                        &
$-$7.79                        &
2.25                           &
1.87                           \\
IRAS05112$-$6755               &
$-$5.96                        &
$-$6.54                        &
3.22                           &
0.58                           \\
IRAS05113$-$6739               &
$-$5.72                        &
$-$6.63                        &
2.96                           &
0.91                           \\
IRAS05128$-$6455               &
$-$6.03                        &
$-$7.94                        &
2.00                           &
1.91                           \\
IRAS05190$-$6748               &
$-$5.72                        &
$-$5.39                        &
3.40                           &
$-$0.33                        \\
IRAS05289$-$6617               &
$-$4.45                        &
                               &
                               &
                               \\
                               &
$-$3.84                        &
                               &
                               &
                               \\
IRAS05295$-$7121               &
$-$5.20                        &
$-$6.74                        &
2.40                           &
1.54                           \\
IRAS05300$-$6651               &
$-$5.54                        &
$-$6.29                        &
3.10                           &
0.75                           \\
IRAS05348$-$7024               &
$-$6.20                        &
$-$6.89                        &
2.90                           &
0.69                           \\
                               &
$-$5.63                        &
$-$5.79                        &
3.40                           &
0.16                           \\
IRAS05360$-$6648               &
$-$5.31                        &
$-$6.19                        &
2.80                           &
0.88                           \\
IRAS05506$-$7053               &
$-$5.54                        &
$-$4.89                        &
3.30                           &
1.35                           \\
IRAS05568$-$6753               &
$-$5.40                        &
                               &
                               &
                               \\
SHV0454030$-$675031            &
$-$3.53                        &
$-$6.64                        &
                               &
3.11                           \\
SHV0500193$-$681706            &
$-$4.79                        &
$-$7.79                        &
1.20                           &
3.00                           \\
                               &
$-$5.63                        &
$-$8.24                        &
1.25                           &
2.61                           \\
SHV0500233$-$682914            &
$-$5.54                        &
$-$8.39                        &
1.55                           &
2.85                           \\
SHV0502469$-$692418            &
$-$4.46                        &
$-$7.68                        &
0.51                           &
3.22                           \\
SHV0521050$-$690415            &
$-$6.30                        &
$-$9.26                        &
0.83                           &
2.96                           \\
SHV0522118$-$702517            &
$-$4.52                        &
$-$7.49                        &
1.30                           &
2.97                           \\
SHV0526001$-$701142            &
$-$4.92                        &
$-$8.04                        &
1.30                           &
3.12                           \\
SHV0535442$-$702433            &
$-$4.67                        &
$-$7.84                        &
0.35                           &
3.17                           \\
TRM45                          &
$-$4.86                        &
$-$6.94                        &
2.05                           &
2.08                           \\
TRM72                          &
$-$5.45                        &
$-$7.53                        &
2.36                           &
2.08                           \\
TRM88                          &
$-$5.59                        &
$-$8.25                        &
1.69                           &
2.66                           \\
WBP14                          &
$-$4.73                        &
$-$7.94                        &
1.05                           &
3.21                           \\
\hline
\end{tabular}
\end{table}

\section{Discussion}

\subsection{Luminosities}

Luminosities derived from the model fits are listed for M-type and carbon
stars in Tables 6 \& 7, respectively, together with $(K-L)$ colours and
bolometric corrections $BC_{\rm K}$ relative to the K-band magnitudes. In
Fig.\ 7 these bolometric corrections are plotted against the $(K-L)$ colours.
The carbon stars define a rather tight relation of the form
\begin{equation}
BC_{\rm K} = 3 - (K-L)^3/12
\end{equation}
valid for $0<(K-L)<4$ mag. Bolometric magnitudes can be estimated for obscured
stars in this colour interval with an accuracy of a few $\times0.1$ mag. If
the scatter in the data is due to limited accuracies of the photometry and/or
model fitting, then luminosities can be estimated with help of the mean
relationship to an accuray of perhaps only $\sim0.1$ mag. M-type stars obey
approximately the same relation between $BC_{\rm K}$ and $(K-L)$ as carbon
stars do, although the scatter in the M-type stars seems larger and they
appear to have $BC_{\rm K}$ smaller by $\sim0.19\pm0.09$ mag for $(K-L)<1$
mag.

%
%
\begin{figure}[tb]
\centerline{\psfig{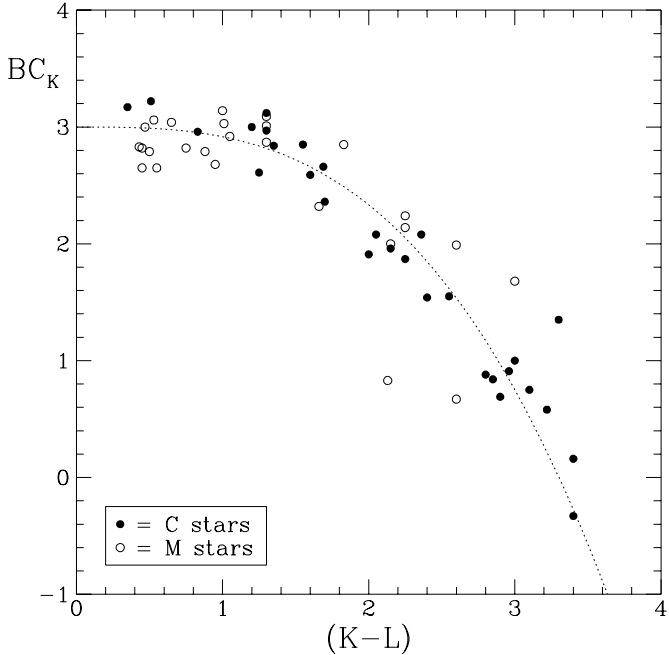}}
\caption[]{Bolometric corrections $BC_{\rm K}$ relative to the K-band
magnitudes, versus $(K-L)$ colours. The relation is virtually identical for
carbon and M-type stars, and is well represented by the function given in Eq.\
(3) (dotted line).}
\end{figure}

%
%
\begin{figure}[tb]
\centerline{\psfig{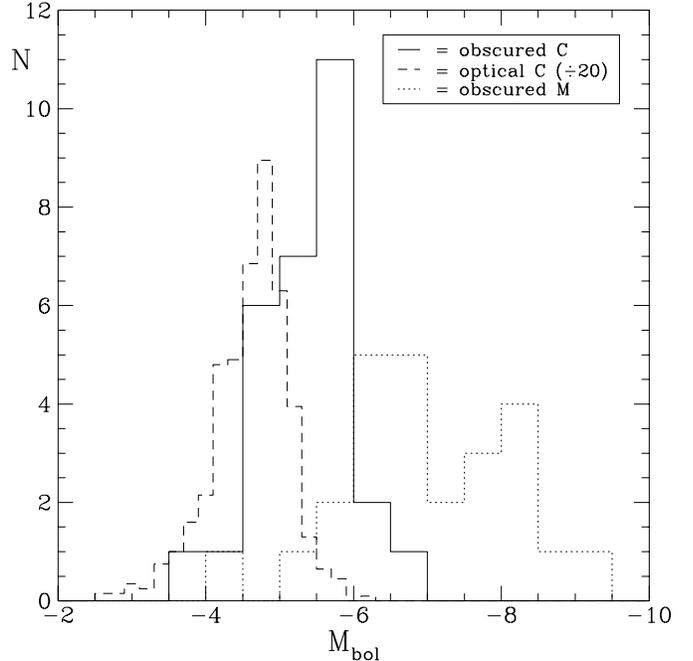}}
\caption[]{Bolometric luminosity distribution of carbon and M-type stars with
$\dot{M} \geq 10^{-7}$ M$_\odot$ yr$^{-1}$, compared to the optically bright
carbon stars in the LMC (divided by 20) from Costa \& Frogel (1996). The
luminosity distribution of dust-enshrouded carbon stars peaks at a luminosity
higher than optically bright carbon stars, but lower than dust-enshrouded
M-type stars. RSGs and AGB stars are separated by a gap in the luminosity
distribution around $M_{\rm bol}\sim-7.5$ mag.}
\end{figure}

The bolometric luminosity distributions of the carbon and M-type stars are
plotted in Fig.\ 8. The distributions resemble those presented in van Loon et
al.\ (1999a). Only stars with $\dot{M} \geq 10^{-7}$ M$_\odot$ yr$^{-1}$ are
plotted, together with the optically bright carbon stars in the LMC from Costa
\& Frogel (1996). Optically bright carbon stars, dust-enshrouded carbon stars,
and dust-enshrouded M-type AGB stars and RSGs occupy distinct luminosity
ranges with little overlap: optically bright carbon stars populate the AGB
mostly between $M_{\rm bol} = -4$ and $-5$ mag, whilst most dust-enshrouded
carbon stars have luminosities between $M_{\rm bol} = -5$ and $-6$ mag. M-type
AGB stars are mainly found between $M_{\rm bol} = -6$ and $-7$ mag, and
(M-type) RSGs peak around $M_{\rm bol} = -8$ mag.

Carbon stars result from the dredge-up of carbon during thermal pulses. The
fact that the dust-enshrouded carbon stars are more luminous than the
optically bright carbon stars suggests that (i) mass-loss rates increase
during the thermal-pulsing AGB (TP-AGB) rendering carbon stars optically
invisible before they leave the AGB and/or (ii) carbon stars with more massive
cores spend a larger fraction of their TP-AGB lifetime being obscured by a
massive CSE than carbon stars with less massive cores. If (i) is true then
dust-enshrouded carbon stars are closer to the tip of  the AGB than are their
optically visible counterparts, in which case their luminosity distribution
(Fig.\ 8) will be representative of the progenitors of Planetary Nebulae. This
is also valid for the M-type AGB stars that do not seem to evolve much in
luminosity once they have become dust-enshrouded: if their luminosities would
still increase by more than a factor of two then many more dust-enshrouded
M-type AGB stars would be expected with $M_{\rm bol}<-6$ mag.

Among the dust-enshrouded AGB stars, the luminosity distribution of M-type
stars peaks at $M_{\rm bol}\sim-6.5$ mag which is significantly more luminous
than the peak of the carbon star luminosity distribution at $M_{\rm
bol}\sim-5.7$ mag. If, as suggested above, luminosities of dust-enshrouded AGB
stars can be directly translated into progenitor masses then the division of
$M_{\rm bol}\sim-6.0$ mag between the dust-enshrouded M-type and carbon star
luminosity distributions translates to a Zero Age Main Sequence (ZAMS) mass of
$M_{\rm div}\sim4$ M$_\odot$ (Wood 1998). Stars with $M_{\rm ZAMS} < M_{\rm
div}$ become carbon stars on the TP-AGB. Our sample of dust-enshrouded carbon
stars contains stars as faint as $M_{\rm bol}\sim-4$ mag, implying a lower
limit to the progenitor mass of carbon stars between 1 and 2 M$_\odot$ (Wood
1998). This lower mass limit should be regarded as tentative, because the way
our sample was selected will result in faint dust-enshrouded stars being
under-represented. It is in good agreement, though, with the 1.5 M$_\odot$
lower limit to carbon star formation as estimated by Groenewegen \& de Jong
(1993).

Stars with $M_{\rm ZAMS} > M_{\rm div}$ stay oxygen rich on the TP-AGB. This
is expected if Hot Bottom Burning (HBB: Iben 1981; Iben \& Renzini 1983; Wood
et al.\ 1983) is effective: in sufficiently massive stars the pressure and
temperature at the bottom of the convective mantle are high enough for carbon
to be processed into (mainly) nitrogen, thus preventing the formation of a
carbon star. The fact that dust-enshrouded carbon stars more luminous than
$M_{\rm bol}=-6$ mag are not entirely absent in the LMC means that HBB cannot
be effective during the entire duration of the TP-AGB. HBB may shut off near
the very end of the TP-AGB when mass loss has reduced the mass of the mantle
to below a critical value needed to maintain the high pressure and temperature
required for HBB to occur (Boothroyd \& Sackmann 1992). Dredge-up of carbon by
another thermal pulse may then cause the photospheric carbon-to-oxygen ratio
to exceed unity (Frost et al.\ 1998; Marigo et al.\ 1998). With 3 out of 13
dust-enshrouded AGB stars between $M_{\rm bol}=-6$ and $-7$ mag being carbon
stars, this means that massive AGB stars (typically $M_{\rm ZAMS}\sim5$
M$_\odot$) experience about four to five thermal pulses during the
dust-enshrouded phase. AGB stars with $M_{\rm ZAMS}\sim5$ M$_\odot$ have
interpulse periods of $\sim1\times10^4$ yr (Vassiliadis \& Wood 1993),
yielding lifetimes for these dust-enshrouded AGB stars of $\sim5\times10^4$
yr.

\subsection{Mass-loss rates}

%
%
\begin{figure}[tb]
\centerline{\psfig{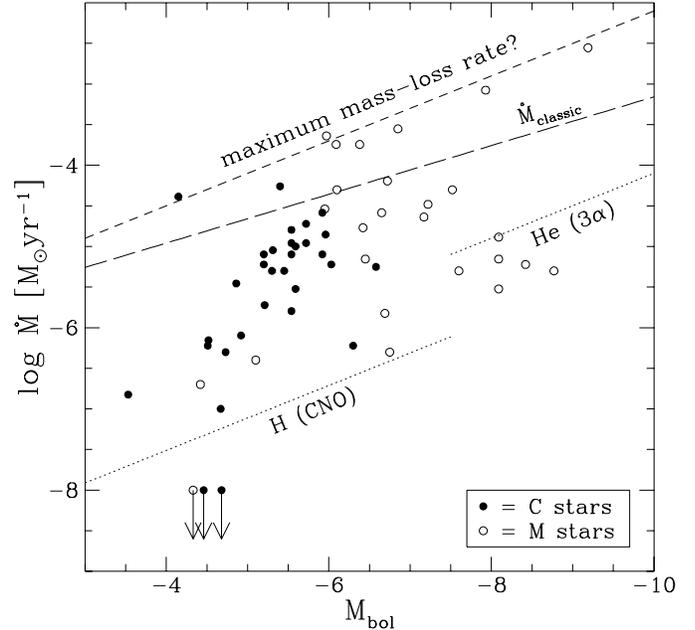}}
\caption[]{Mass-loss rates versus bolometric luminosities for carbon and
M-type stars. The distribution of stars suggests a maximum limit to $\dot{M}$
which is proportional to $L$ (short-dashed line) and higher than the classical
limit $\dot{M}_{\rm classic} = L (v_{\rm exp} c)^{-1} \propto L^{0.75}$
(long-dashed line). The rates at which mass is burned by the CNO (AGB stars)
and 3$\alpha$ (RSGs) cycles are also proportional to $L$ (dotted lines).}
\end{figure}

In Fig.\ 9 the mass-loss rates are plotted versus the bolometric magnitudes
for the stars in our sample. Before discussing the implications for the mass
loss in the AGB and RSG phases, two possible selection effects need to be
addressed.

First, faint stars with high mass-loss rates may have escaped selection due to
the difficulty to detect their faint, obscured near-IR counterpart. Indeed,
the two carbon stars IRAS05289$-$6617 with $M_{\rm bol} = -4.15$ mag and
$\dot{M} = 4.1\times10^{-5}$ M$_\odot$ yr$^{-1}$ and IRAS05568$-$6753 with
$M_{\rm bol} = -5.40$ mag and $\dot{M} = 5.5\times10^{-5}$ M$_\odot$ yr$^{-1}$
were only included because in both cases the IRAS point source had been
erroneously identified with an M-type field star. The K-band magnitude of
their real near-IR counterparts is expected to be fainter than $K \sim 19$
mag, which is beyond any of the near-IR searches done so far. Hence more such
heavily obscured, bolometrically faint stars are expected to exist (see also
paper III and Loup et al.\ 1999). From the luminosity distributions (Fig.\ 8)
these are expected to be carbon stars. A similarly obscured object (S13) was
found in the Small Magellanic Cloud by Groenewegen \& Blommaert (1998).

Second, stars with small IR excesses have been largely excluded from our
sample. This may have resulted in the absence of stars with $\dot{M} \lsim
10^{-7}$ M$_\odot$ yr$^{-1}$ around $M_{\rm bol} \sim -4$ mag to stars with
$\dot{M} \lsim 10^{-6}$ M$_\odot$ yr$^{-1}$ around $M_{\rm bol} \sim -9$ mag.
It would be interesting to investigate whether this region in the $\dot{M}$
versus $M_{\rm bol}$ diagram is in fact devoid of stars. Gail \& Sedlmayr
(1987) suggested that dynamical considerations exclude dust-driven winds with
$\dot{M}\lsim$ few $\times10^{-6}$ M$_\odot$ yr$^{-1}$ (at $L\sim$ few
$\times10^3$ L$_\odot$), which is reproduced in recent model calculations by
Schr\"{o}der et al.\ (1999). The values for $\dot{M}$ that we find for faint
stars of $L\sim$ few $\times10^3$ L$_\odot$ are more than an order of
magnitude lower than the lower limit predicted by Gail \& Sedlmayr (1987).
These stars provide direct evidence for the existence of a wind-driving
mechanism other than purely radiation pressure on dust.

On average mass-loss rates appear to be higher for the M-type stars than for
the carbon stars in our sample, but this is a result of the higher
luminosities of the M-type stars in our sample rather than the different
chemistry per se (Fig.\ 9). The data suggest a maximum limit to the mass-loss
rate, outlined by a (short-dashed) line of $\dot{M} \propto L$. There does not
seem to be a distinction between the luminosity dependence of the maximum
$\dot{M}$ for AGB stars --- be it carbon or M-type --- and RSGs, despite their
different masses and internal structure. The maximum $\dot{M}$ is higher than
the classical single-scattering limit $\dot{M}_{\rm classic} = L (v_{\rm exp}
c)^{-1} \propto L^{0.75}$ (Jura 1984) by a factor $\sim2$ at $L = 10^3$
L$_\odot$ up to a factor $\sim10$ at $L = 5\times10^5$ L$_\odot$. This is
expected to happen when multiple scattering of photons in the dusty CSE
becomes important (Gail \& Sedlmayr 1986). The six M-type stars closest to the
$\dot{M}$ limit are indeed also the stars with the largest optical depth at 1
$\mu$m among the M-type stars, averaging $\tau_{\rm rad}$(1 $\mu$m)$ = 9\pm2$.
The same is true for the two carbon stars near the $\dot{M}$ limit,
IRAS05289$-$6617 and IRAS05568$-$6753, that have the largest optical depth
among the carbon stars, with $\tau_{\rm rad}$(1 $\mu$m)$ = 76$ and 56,
respectively.

%
%
\begin{figure}[tb]
\centerline{\psfig{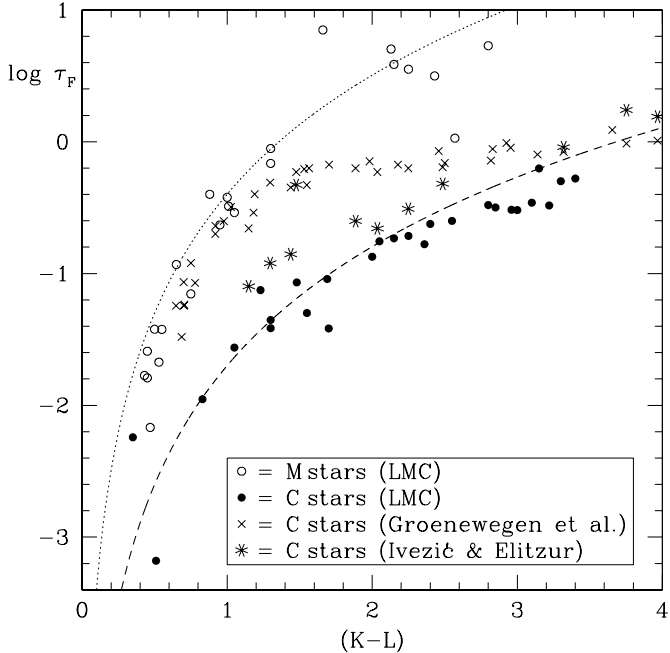}}
\caption[]{Flux-weighted optical depth $\tau_{\rm F}$ versus $(K-L)$ colour.
The correlation for M-type stars in the LMC (open circles) is approximated by
the dotted line, whereas the correlation for carbon stars in the LMC (solid
circles) is approximated by the dashed line. Also plotted are galactic carbon
stars, with $\tau_{\rm F}$ values from Groenewegen et al.\ (1998) (crosses)
and Ivezi\'{c} \& Elitzur (1995) (stars).}
\end{figure}

The factor $\dot{M}/\dot{M}_{\rm classic}$ can be identified with the
flux-weighted optical depth $\tau_{\rm F}$ (cf.\ Ivezi\'{c} \& Elitzur 1995;
Groenewegen et al.\ 1998, and references therein). In Fig.\ 10 the values of
$\tau_{\rm F} = \dot{M}/\dot{M}_{\rm classic}$ for the M-type stars (open
circles) and carbon stars (solid circles) in our ISO sample in the LMC are
plotted against their $(K-L)$ colours. Not surprisingly, these two quantities
are well correlated and can be reasonably well approximated by $\tau_{\rm F}
\sim0.4\times (K-L)^3$ for the M-type stars (dotted line) and $\tau_{\rm F}
\sim0.02\times (K-L)^3$ for the carbon stars (dashed line). From Fig.\ 10 it
can be seen that multiple scattering already becomes important ($\tau_{\rm
F}\gsim1$) for M-type stars with $(K-L){\gsim}1.5$ mag, whereas for carbon
stars it only becomes important for $(K-L){\gsim}4$ mag. The fact that the two
LMC carbon stars with $\tau_{\rm F}>1$ had not been identified in the near-IR
is consistent with their expected $(K-L)\gg4$ mag on the basis of Fig.\ 10.
The galactic carbon stars from Groenewegen et al.\ (1998) are plotted for
comparison (crosses). These are real flux-weighted optical depths, i.e.\ they
were not calculated from the mass-loss rate, luminosity and expansion
velocity. The flux-weighted optical depths of the galactic carbon stars from
Ivezi\'{c} \& Elitzur (1995) are plotted too (stars). For both galactic
samples we use the $(K-L)$ colours of the model fits from Groenewegen et al.,
after transformation onto the SAAO system (Carter 1990). The Ivezi\'{c} \&
Elitzur values of $\tau_{\rm F}$ are in better, but not exact, agreement with
the carbon stars in the LMC. The reason for the discrepancy between the
galactic carbon stars and the carbon stars in the LMC, and between the work of
Groenewegen et al.\ and Ivezi\'{c} \& Elitzur is likely to be related to the
different optical properties assumed for the dust by the various authors.

To appreciate the importance of the mass-loss rate for the evolution of a
star, $\dot{M}$ should be compared with the rate at which mass is consumed by
nuclear burning, $\dot{M}_{\rm nuc}$, and which determines the evolutionary
timescale in the absence of severe mass loss. In AGB stars the energy is
mainly produced by the CNO cycle in the H-burning shell, releasing
$\sim6.1\times10^{18}$ erg g$^{-1}$, whilst in RSGs the 3$\alpha$ cycle in the
He-burning core is responsible for the stellar luminosity, releasing
$\sim5.9\times10^{17}$ erg g$^{-1}$ (Kippenhahn \& Weigert 1990). The lines of
$\dot{M}_{\rm nuc}$ are proportional to $L$ (dotted in Fig.\ 9). All AGB stars
in our sample with measurable mid-IR emission from circumstellar dust have
$\dot{M}>\dot{M}_{\rm nuc}$, and hence their AGB evolution will be terminated
by the loss of the stellar mantle before the core can grow significantly.
Indeed, the extreme AGB stars with $\tau_{\rm F}>1$ have lifetimes of only of
the order of $10^4$ yr, which is comparable to the interpulse timescale
(Vassiliadis \& Wood 1993). Most RSGs ($M_{\rm bol}<-7.5$ mag), however,
experience $\dot{M}\sim\dot{M}_{\rm nuc}$. This means that they cannot lose a
large fraction of their stellar mantles before presumably ending as
supernovae. WOH G64 with $M_{\rm bol} = -9.19$ mag and $\dot{M} =
2.8\times10^{-3}$ M$_\odot$ yr$^{-1}$ and probably also IRAS04530$-$6916 with
$M_{\rm bol} = -7.93$ mag and $\dot{M} = 8.4\times10^{-4}$ M$_\odot$ yr$^{-1}$
may be the only RSGs in our sample that have $\dot{M}{\gg}\dot{M}_{\rm nuc}$.
At such high mass-loss rates these two stars have lifetimes of only
$\sim2\times10^4$ yr.

The empty area in the ($M_{\rm bol}$,$\dot{M}$) diagram between the two RSGs
WOH G64 and IRAS04530$-$6916 with $\log \dot{M}=-2.6$ and $-3.1$,
respectively, and the six RSGs with about two orders of magnitude lower
mass-loss rates is hard to explain in terms of selection effects and may
therefore be a real gap in the $\dot{M}$ distribution of RSGs. From these
numbers we estimate that RSGs spend about 25\% of their RSG lifetime in the
intense mass-loss phase at $\dot{M}\sim10^{-3}$ M$_\odot$ yr$^{-1}$ and 75\%
in a moderate mass-loss phase at $\dot{M}\sim10^{-5}$ M$_\odot$ yr$^{-1}$. A
similar gap of about an order of magnitude in $\dot{M}$ may be present in the
$\dot{M}$ distribution of AGB stars, although for stars fainter than $M_{\rm
bol}\sim-6$ mag this is hard to prove due to the selection effects mentioned
before. If true, however, this means for the AGB stars that most ($\sim80$\%)
of the time on the TP-AGB is spent in relative quiescence, at high but not
extraordinarily high $\dot{M}$, with only short phases of intense mass loss
(see also Vassiliadis \& Wood 1993). Such a ``burst'' may be episodic, or it
may be the final shedding of the stellar mantle before the star leaves the
AGB. According to recent evolutionary calculations by Schr\"{o}der et al.\
(1998, 1999), which take into account the characteristics of dust-induced
mass-loss from pulsating AGB stars, the duration of the high mass-loss phases
strongly depends on the initial stellar mass.

The two brightest carbon stars in the sample, IRAS04496$-$6958 with $M_{\rm
bol}=-6.6$ and SHV0521050$-$690415 with $M_{\rm bol}=-6.3$ mag have mass-loss
rates that are significantly lower than the mass-loss rates of most of the
oxygen-rich AGB stars at these high luminosities. We speculate that these
stars may exhibit lower mass-loss rates following a (final) thermal pulse
(Schr\"{o}der et al.\ 1998, 1999) that converted them from oxygen-rich AGB
stars into carbon stars after HBB could not be maintained anymore because the
stellar mantles had been strongly diminished by mass loss on the TP-AGB. This
strengthens the explanation of the silicate dust around IRAS04496$-$6958 being
due to a final thermal pulse, as suggested by Trams et al.\ 1999a.

%
%
\begin{figure}[tb]
\centerline{\psfig{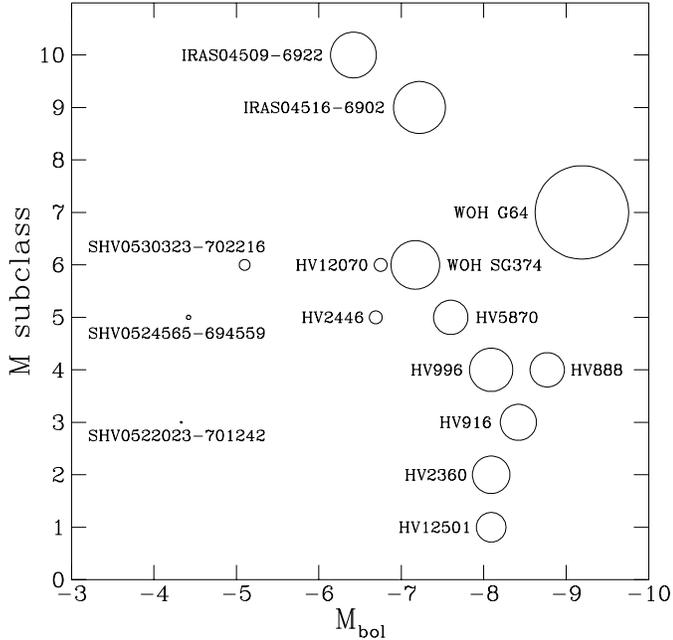}}
\caption[]{Spectral subclasses versus bolometric luminosities for the M-type
stars with known spectral type. The size of the symbol increases with larger
$\dot{M}$. Stars of later spectral type (cooler $T_{\rm eff}$) and higher
luminosity have larger $\dot{M}$.}
\end{figure}

The process of dust formation is extremely sensitive to the temperature (Arndt
et al.\ 1997), and (episodes of) increased mass-loss rates may therefore be
related to lower $T_{\rm eff}$ of the stars. The spectral subclasses of the
M-type stars in our sample for which the spectral type is known are plotted
versus their bolometric magnitudes (Fig.\ 11), with the size of the symbol
increasing with larger $\dot{M}$. The strong luminosity dependence of
$\dot{M}$ is very prominent. However, there also seems to be a temperature
dependence of $\dot{M}$: cooler stars, i.e.\ stars with larger subclasses,
experience larger $\dot{M}$ than warmer stars. Because the luminosity,
effective temperature and surface gravity are inter-related (via the stellar
radius) it is not yet clear to what extent the dependence of $\dot{M}$ on $L$
and $T_{\rm eff}$ reflects a dependence of $\dot{M}$ on $\log g$.

\section{Summary}

We derived mass-loss rates and luminosities for a sample of dust-enshrouded
oxygen- and carbon-rich AGB stars as well as a few optically bright AGB stars
and red supergiants, all of which have been observed with ISO, using a
radiation transfer code to model the spectral energy distributions at IR
wavelengths.

The luminosity distribution of dust-enshrouded carbon stars peaks at $M_{\rm
bol}\sim-5.7$ mag, which is in between the peak of the luminosity distribution
of optically bright carbon stars ($M_{\rm bol}\sim-4.8$ mag) and the peak of
the luminosity distribution of dust-enshrouded M-type AGB stars ($M_{\rm
bol}\sim-6.5$ mag). We interpret this as follows: stars with masses of $\sim2$
to 4 M$_\odot$ become dust-enshrouded carbon stars on the thermal pulsing AGB,
whereas more massive AGB stars remain oxygen-rich as a result of Hot Bottom
Burning. Optically bright carbon stars may still evolve in luminosity before
becoming dust-enshrouded, or they may be associated with generally less
massive stars (1 to 2 M$_\odot$) than the stars that become dust-enshrouded
carbon stars. The existence of a few luminous carbon stars may be explained by
HBB switching off near the very tip of the AGB and the occurrence of a final
thermal pulse thereafter.

Mass-loss rates of both AGB stars and RSGs are found to increase with
increasing luminosity and decreasing effective temperature, and are typically
$\dot{M}\sim10^{-6}$ to $10^{-4}$ M$_\odot$ yr$^{-1}$ on the TP-AGB and
$\dot{M}\sim10^{-5}$ to $10^{-3}$ M$_\odot$ yr$^{-1}$ for RSGs. Mass-loss
rates in some cases exceed the limit set by the rate at which the available
photon momentum can be transferred onto the gas and dust by means of single
scattering, providing evidence for multiple scattering to be effective in
driving the outflow. Such intense mass loss only happens during distinct
(possibly recurrent) phases in the evolution of AGB stars and RSGs that last
typically 20 to 25\% of the dust-enshrouded lifetime. It is only during these
episodes of intensified mass loss that the mass-loss rate of RSGs exceeds the
mass-consumption rate due to nuclear burning, whereas the mass-loss rate of
dust-enshrouded AGB stars always exceeds the mass-consumption rate due to
nuclear burning.

\acknowledgements{We would like to thank Dr.\ Carlos L\'{a}zaro for making his
near-IR spectra of carbon stars available in electronic form, Prof.Dr.\ Teije
de Jong and Jeroen Bouwman for valuable suggestions, the referee Dr.\ Jan
Martin Winters for his excellent remarks that helped improve the presentation
and discussion of the results, and Joana Oliveira for everything. We made use
of the Astronomical Data Center, operated at NASA GSFC. A major part of this
research was performed when JvL was at the University of Amsterdam, and
supported by NWO Pionier Grant 600-78-333. AdK also acknowledges support from
NWO Spinoza Grant 08-0 to E.P.J.\ van den Heuvel. JvL adds: Anjo, agora sou
todo teu!}

\end{document}